\documentclass[runningheads]{llncs}
\usepackage{graphicx}
\usepackage{subfigure} 
\begin{document}
\title{HSEarch: semantic search system for workplace accident reports}
\author{Emrah Inan\inst{1} \and
Paul Thompson\inst{1} \and
Tim Yates\inst{2}
\and Sophia Ananiadou\inst{1}\thanks{Corresponding author.}}
\authorrunning{E. Inan et al.}
\institute{
\textsuperscript{1}National Centre for Text Mining, University of Manchester, Manchester, UK \\
\textsuperscript{2}Health \& Safety Executive, HSE Science \& Research Centre, Buxton, UK
\email{(paul.thompson,emrah.inan,sophia.ananiadou)@manchester.ac.uk, tim.yates@hse.gov.uk}}
\maketitle              % typeset the header of the contribution
\begin{abstract}
  Semantic search engines, which integrate the output of text mining (TM) methods, can significantly increase the ease and efficiency of finding relevant documents and locating important information within them. We present a novel search engine for the construction industry, HSEarch (http://www.nactem.ac.uk/hse/), which uses TM methods to provide semantically-enhanced, faceted search over a repository of workplace accident reports. Compared to previous TM-driven search engines for the construction industry, HSEarch provides a more interactive means for users to explore the contents of the repository, to review documents more systematically and to locate relevant knowledge within them.

\end{abstract}
\section{Introduction}
\label{intro}
Ensuring safety in new construction projects requires an exploration of documents describing potential hazards and mitigations from previous projects that share similar sets of attributes. Text mining (TM) methods have been used in construction document retrieval systems to expand queries with additional semantically-related terms (e.g. \cite{kim2019accident}), to retrieve semantically similar documents (e.g., \cite{zou2017retrieving}) and to recognise concepts automatically (e.g., \cite{gao2017bimtag}). In other domains, semantic search systems allowing filtering of results based on various \textit{facets} of semantic content have been effective (e.g., \cite{thompson2016text}). 

In this paper, we present a novel search system for construction-related documents, HSEarch, which facilitates search over 3000 Reporting of Injuries, Diseases and Dangerous Occurrences Regulations (RIDDOR) workplace accident reports from the archive of the Health and Safety Executive (HSE). The system integrates standard keyword-based search with state-of-the-art TM methods to provide faceted search refinement at different levels of granularity, while automatic summarisation increases the efficiency of scanning longer documents for potential relevancy.  Compared to other construction industry search systems, HSEearch provides a more interactive and flexible environment for efficient exploration and filtering of workplace accident reports from multiple perspectives.    

\section{Related Work}
\label{sectionRelWork}
Previous TM-based studies have aimed to ease the burden of retrieving and exploring construction documents, e.g., a search system over computer-aided design (CAD) documents uses similarity between text extracted from the documents and the input query as the basis for retrieval \cite{hsu2013content}. Automatic document classification approaches have used pre-defined topics from a construction information classification system \cite{caldas2003automating} or different categorisations of injuries, incidents or hazards \cite{bertke2012development,taylor2014near,goh2017construction}. In \cite{tixier2016automated}, dictionaries and rules are used to recognise pre-defined injury-related concepts in texts. However, supervised Named Entity Recognition (NER) methods are more flexible, since they learn how to recognise mentions of concepts that never occur in the training data. A corpus of RIDDOR reports \cite{thompson2020semantic}, manually annotated with 6 concept categories (e.g., hazards, consequences, and project attributes) facilitates supervised NER for the construction domain. 

\section{HSEarch}
\label{sectionMethod}
HSEarch was implemented using Elasticsearch\footnote{https://
www.elastic.co/products/elasticsearch}. It can be accessed from a
web-based interface written in Flask\footnote{https://flask.palletsprojects.com/en/1.1.x/}, a microframework for web development.

\begin{figure} [ht]
\centering
  \includegraphics[scale=0.2]{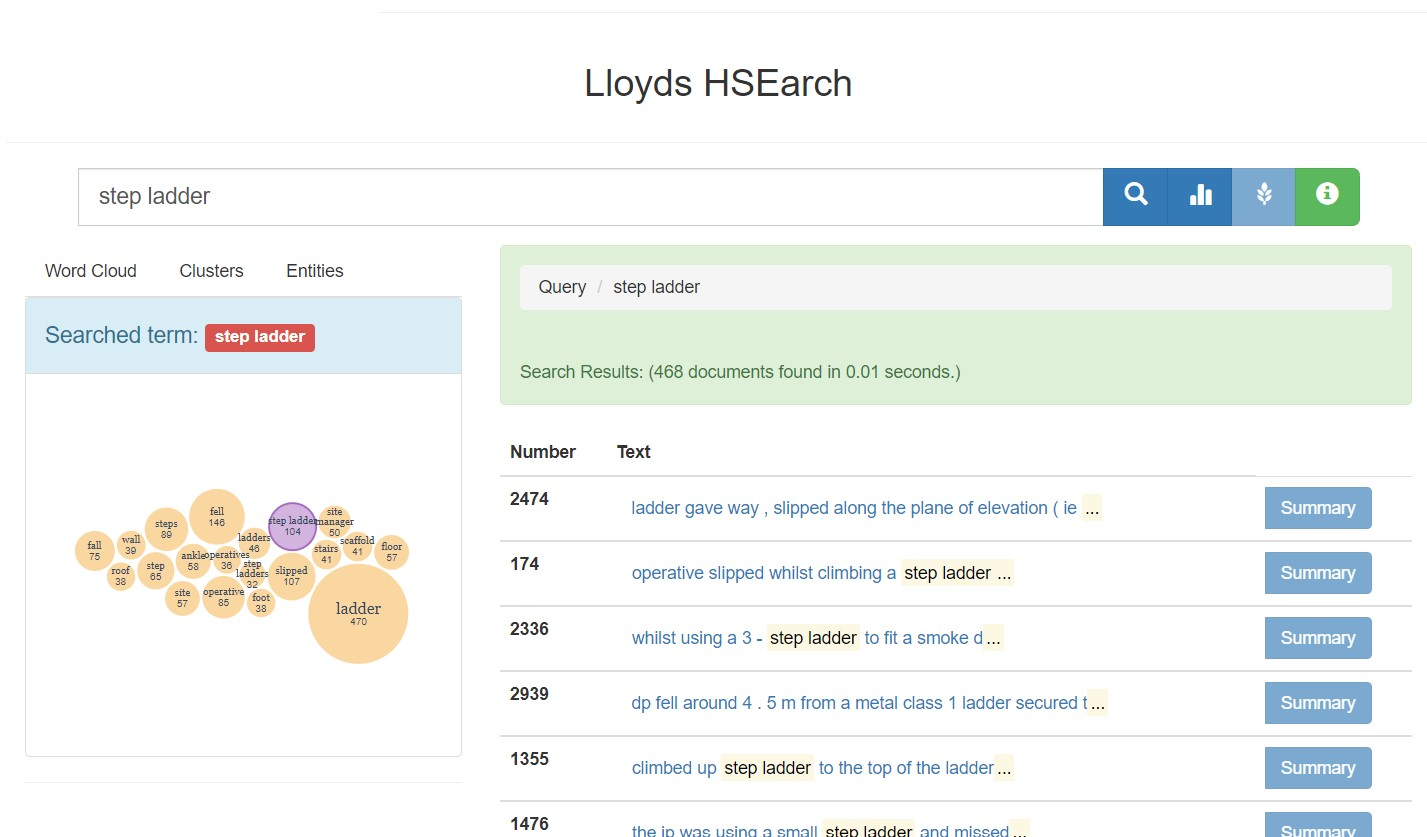}
\caption{HSEarch user interface}
\label{figGraStru}
\end{figure}

Figure 1 illustrates the three main components of the user interface, i.e., a search area (top); main search results pane (right); and content exploration pane (left), consisting of three different tabs (\textit{word cloud}, \textit{clusters} and \textit{entities}), allowing the semantic content of the retrieved documents to be explored/filtered.

\begin{figure}[ht]
\centering
\subfigure[Descriptive clusters]{%
\label{fig:f4:Clusters}%
\includegraphics[scale=0.32]{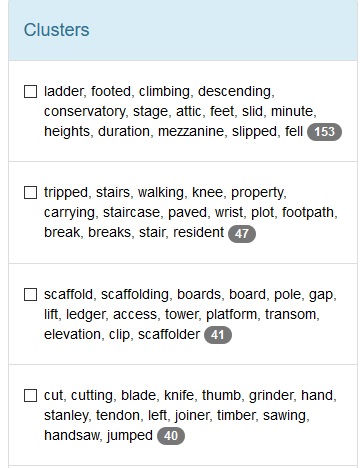}}%
\qquad
\subfigure[Construction activity entities]{%
\label{fig:f4:ConstructionAct}%
\includegraphics[scale=0.32]{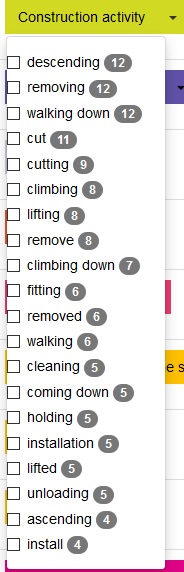}}%
\qquad
\subfigure[Harmful consequence entities]{%
\label{fig:f4:HarmConseq}%
\includegraphics[scale=0.32]{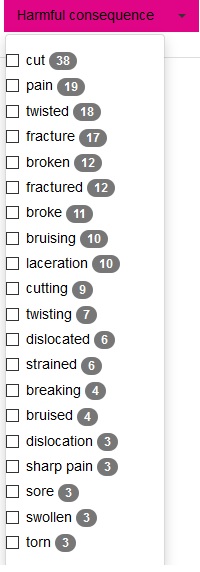}}%
\caption{Clusters and entity categories}
\end{figure}

The left of Figure 1 illustrates the word cloud resulting from a search for \textit{slipped}. The cloud provides a dynamically-generated overview of the content of the retrieved documents, obtained using the widely-used TerMine tool \cite{frantzi2000automatic}, which automatically identifies the most important terms mentioned within a collection of documents. The \textit{Summary} button displays a short summary of longer documents, produced using an entity enriched graph-based method \cite{mihalcea2004textrank} whose nodes including NEs and TerMine-identified terms. We compute similarity between the nodes, leveraging the Word2vec model \cite{mikolov2013distributed} trained using the RIDDOR reports. After obtaining a weighted graph, we employ the PageRank \cite{brin1998anatomy} algorithm to rank the most representative sentences. We then apply Maximum Marginal Relevance (MMR) \cite{carbonell1998use} to the ranked list to ensure that only sentences providing new information are added to the summary.

Figure 2(a) displays some clusters for the search term \textit{slipped}, which provide a high-level overview of the most pertinent topics covered in the retrieved documents.  Clicking a check box next to a cluster will filter the search results to retain only the documents in the selected cluster. We use a recently developed, self-tuned \textit{descriptive clustering} approach \cite{brockmeier2018self}, in which the set of topics is dynamically determined for each new search, and documents are automatically clustered according to these search-specific topics.  While the first three clusters clearly correspond to slipping incidents where a fall took place, the documents in fourth cluster concern equipment slipping from a worker's grasp. 

\begin{figure} [ht]
\centering
  \includegraphics[scale=0.28]{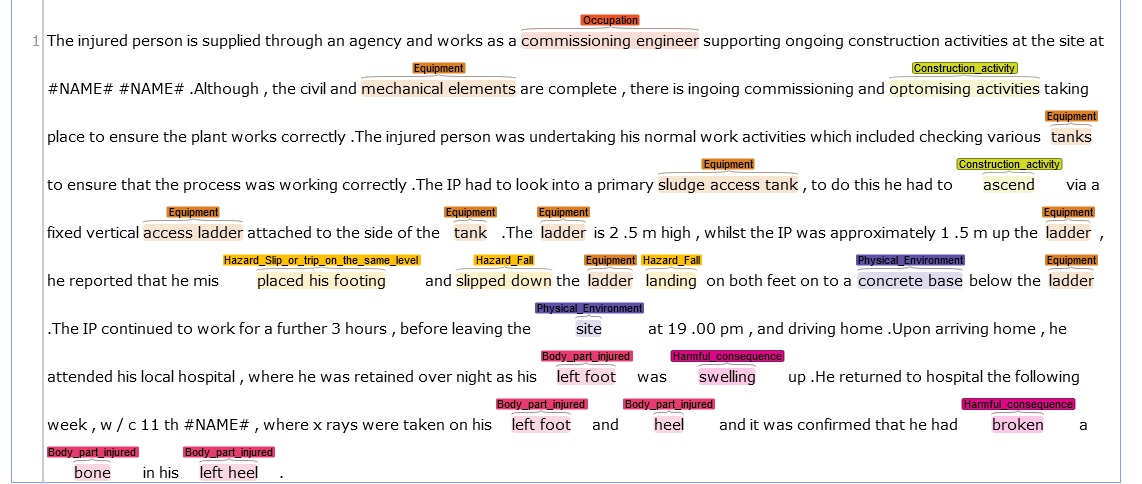}
\caption{Semantically-enhanced full text document view}
\end{figure}

We apply a layered neural model \cite{ju2018neural} to the construction safety corpus \cite{thompson2020semantic}\footnote{Consult cited paper for details of documents and categories annotated} to recognise domain-relevant NEs.  Figure 2(b) shows common NEs mentioned in documents containing the search term \textit{slipped}.  While most activities are concerned with ascending or descending, others, like lifting or unloading items, also carry a risk of slipping. Figure 2(c) shows the most frequent \textit{Harmful consequence} NEs in documents containing the word \textit{slipped}, which vary in severity from bruisings to breaks. This list could be used to prioritise exploration of the most severe consequences. Highlighting and colour-coding of NEs in documents (Figure 3) makes it easy to focus on parts of the text containing different types of important information.

\section{Evaluation}
\label{sectionEval}

The relevance of retrieval results for 20 queries based on various aspects of accidents (e.g., risks, causes and equipment) were evaluated on a scale of 0-2 by 4 domain experts, following the TREC evaluation paradigm \cite{harman2012trec}. 

\begin{table}
\caption{Experimental retrieval results using the system.}\label{tab1}
\begin{tabular}{llllllllllll}
\hline
Enquires &  \multicolumn{2}{l}{ Expert 1} & \multicolumn{2}{l}{ Expert 2}
& \multicolumn{2}{l}{ Expert 3} & \multicolumn{2}{l}{ Expert 4} & \multicolumn{2}{l}{ AVG }\\
Category & nDCG & P@5 & nDCG & P@5 & nDCG & P@5 & nDCG & P@5 & nDCG & P@5  \\
\hline
Word-Based & 0.674 & 0.77 & 0.608 & 0.67 & 0.578 & 0.54 & 0.619 & 0.66 & 0.619 & 0.66 \\
Entity-Based & 0.966 & 0.75 & 0.945 & 0.65 & 0.920 & 0.45 & 0.915 & 0.55 & 0.937 & 0.6 \\
\hline
\end{tabular}
\end{table}

Table 1 compares experimental results for entity-indexed texts with traditional word-based indexing in Elasticsearch, using Trectools \cite{palotti2019}. We used Fleiss' kappa (overall score 0.99) to verify the correlation between pairs of relevance assessments, and Kendall's tau (0.768) to verify system correlation between word and entity-based indexing. Although P@5 (precision for the first 5 results) is generally similar for both word and entity-based indexing, nDCG (normalised discounted cumulative gain) is significantly higher for entity-based indexing, showing that this method results in better ranked results. Furthermore, we carried out an initial usability study of the HSEarch system, in which we evaluated whether the semantically enhanced document view can be useful within a given user scenario, in terms of being able to capture the prevalence of a risk category. If a domain expert wants to find an answer to the query ``How many cuts are caused by a Stanley knife blade?", then the fact that the search system identifies ``Stanley knife blade" as an NE of type ``Equipment" allows the user to easily filter documents that mention this NE and explore what is being said about it.

%This paper has described \textit{HSEarch}, a novel, interactive and faceted semantic search system integrating TM methods to support efficient %discovery about workplace accidents in the construction industry. As future work, we will augment the system to allow search over other documents %generated during the construction project lifecycle. The varying nature of these reports may also require the recognition of additional NE types %to allow full use to be made of their content, and hence we also intend to extend the size of the annotated dataset. This larger dataset will also %allow us to improve the quality of our summarisation method, by applying the method described in \cite{zerva2020cited}, which leverages %pre-trained encoders in combination with different neural networks.

\bibliographystyle{splncs04}
\bibliography{samplepaper}
\end{document}